# Charge Transport Modeling of CdSe/ZnS core/shell Quantum Nanorod Light-Emitting Diodes

A.G. Melkonyan[1], G.A. Mantashian[1], D.B. Hayrapetyan[1,2]

[1]Quantum Materials and Nanophotonics Laboratory, A.B. Nalbandyan Institute of Chemical Physics, 5/2 Paruyr Sevak St., Yerevan 0014, Armenia

[2]Department of General Physics and Quantum Nanostructures, Russian-Armenian University, 123 Hovsep Emin Str., Yerevan 0051, Armenia

*Correspondence: david.hayrapetyan@rau.am

**Abstract**: In this study, we investigate the electronic structure, charge transport dynamics, and optical properties of a quantum dot light-emitting diode (QD-LED) featuring a double nanorod (NR) emission layer composed of CdSe/ZnS core/shell structures. Utilizing a rigorous self-consistent numerical approach, we solve the coupled Schrödinger-Poisson equations iteratively to obtain accurate wave functions, energy levels, and potential profiles under varying external bias voltages. Detailed analyses reveal voltage-dependent electron localization dynamics, demonstrating a systematic transition of electrons between distinct NR regions via quantum tunneling. Charge density and electrostatic potential distributions are modeled comprehensively, employing the asymmetric Erlang distribution to characterize interface effects. By calculating current-voltage (I-V) characteristics and photoluminescence spectra, we demonstrate that external voltage serves as a robust tuning parameter for modulating emission energies and intensities, underscoring the potential of these NR-LED systems for tunable optoelectronic and photonic applications.

## 1. Introduction

Quantum dots (QDs) are nanostructures that exist in the range of several to dozens of nanometers and possess the property of quantum confinement which alters their electronic and optical properties compared to their bulk counterparts. In contrast to the bulk material with continuous energy levels, the QDs exhibit discrete quantized energy levels, resulting in very sharp absorption and emission spectra. Their optical band gap can be tuned to be designed to have specific properties. QDs have unique optical properties, including high quantum yield, excellent photobleaching resistance, broad absorption, and narrow emission spectra [1-5]. These features make QDs highly versatile in their applications, including optoelectronics, biotechnology, sensing, solar cells, LEDs, quantum computing, photodetectors, and medicine [6-9].

A distinct advantage over the traditional spherical QDs, is provided by core/shell nanorods a rapidly developing class of quantum-confined nanoparticles. Their core/shell structure offers an additional degree of tunability. Additionally, their rod-like structure enables them to exhibit linearly polarized absorption and emission a larger surface area and more efficient electron-hole charge separation, leading to more efficient electron injection, which is important in optoelectronic devices. Compared to spherical QDs, nanorods also exhibit improved optical absorption due to their larger optical cross-section. This property is very advantageous for nanorods in various applications including gas sensors, optoelectronics, dye-sensitized solar cells, and biosensors [10-23]. Another important advantage of nanorods is that they can promote faster radiative degradation processes and slower photobleaching, making them more stable and effective for long-term use making them ideal for use in advanced nanophotonic devices such as nanocrystal-based lasers [24,25].

In the context of LEDs, nanorods (NRs) offer a number of promising attributes. Nanorod-based LEDs (NR-LEDs) exhibit a wide range of emission spectra that can be precisely tuned by altering the diameter or width of the nanorods, enabling not only visible but also near-infrared (NIR) emission. This tunability, together with a significant Stokes shift and improved light outcoupling, makes NR-LEDs versatile for next-generation photonic applications. In particular, by engineering nanorods for NIR emission, NR-LEDs become highly suitable for specialized applications such as optical communication [26,27], biomedical imaging[28], night vision[29], and remote sensing[30], in addition to their established potential for advanced lighting and display technologies. Unlike spherical QDs, which experience a large drop in quantum yield (QY) when converted from a solution state to tightly packed thin films, nanorods are more effective at suppressing nonradiative energy transfer in densely packed environments due to their elongated geometry, which increases emitter separation and reduces dipole–dipole coupling, thus limiting Förster-type energy losses [31]. This suppression results in higher QY, especially for solid-state applications, which is an important parameter in determining the efficiency of light emission in LEDs. Furthermore, nanorods have unique anisotropic optical features that allow for more precise control of light emission directionality, making them particularly appealing for applications that need polarized light emission. High QY, combined with improved stability and optical characteristics unique to nanorods, make NR-LEDs an excellent contender in the design of highly efficient and tunable light-emitting devices [32-35].

Achieving well-aligned NR arrays is critical for maximizing light outcoupling, emission uniformity, and carrier injection efficiency. A variety of scalable strategies have been developed: electrophoretic deposition to vertically orient CdSe/CdS NRs by applying a bias across colloidal suspensions [31], dielectrophoretic assembly using alternating electric fields to align colloidal and epitaxial NRs between patterned electrodes [36-38], template-assisted patterning via nanosphere lithography and nanoimprint to precisely position InGaN/GaN and CdSe/CdS NRs [39,40], solution-phase self-assembly with ligand and solvent engineering to order colloidal NRs during spin-coating or drop-casting [41,42], and epitaxial growth methods (PA-MBE, MOCVD, VLS) that exploit crystallographic orientation to produce vertical NR arrays with monocrystalline quality [40,42].

In order to simulate the properties in NR-LED devices it is necessary to consider the charge transport processes in them. The charge transport model for NR-LEDs includes tunneling, injection and drift-diffusion currents, to accurately describe charge carrier dynamics. In this model, charge injection and tunneling occur between NRs and the charge transport layers, as well as between NRs within the emissive layer (EML). Drift-diffusion mechanisms govern charge transport in the hole transport layer (HTL) and electron transport layer (ETL), while coupled rate equations account for tunneling and recombination in the NR layers, ensuring precise modeling of conduction and recombination processes [43-45].

Theoretical frameworks play a crucial role in the study of nanostructures due to the complexity of the involved quantum systems and the high cost of experimental methods. In this work, we use single-band k·p theory, which is a widely accepted and computationally efficient method, for calculating the band structures of NRs near band extrema. Unlike other approaches such as density functional theory or tight-binding methods, which are more computationally demanding, the k·p theory offers a simplified but accurate representation of the band-edge states in semiconductor nanostructures [46-49]. Furthermore, to accurately model the electronic states in these quantum-confined systems, we solve the self-consistent Schrödinger-Poisson equations. This approach allows us to obtain an efficient numerical solution to describe the quantum states in confined systems, taking into account both quantum effects and electrostatic interactions inside the NR. In addition, we use the drift-diffusion model to simulate the transport of electrons and holes, as this model has proven successful in modeling semiconductor devices. We provide a comprehensive framework for studying the optical and electronic properties of NRs, by combining these theoretical methods.

## 2. Model

### 2.1. Device Architecture Description

In this section, we explore the architectural layout of the NR-LED device, as depicted in Figure 1. The device has a multilayer structure that consists of a hole injection layer (HIL), hole transport layer (HTL), emission layer (EML), which consists of two columns of NRs which are axially oriented along the field direction, and electron transport layer (ETL) sandwiched between an anode and a cathode (see Figure 1 (a)). An external voltage source supplies electrons to the ETL and injects holes into the HIL/HTL layers. These charge carriers subsequently migrate into the NR-based EML, where radiative recombination occurs. The indium tin oxide (ITO) anode is connected to the HIL, which consists of Poly(3,4-ethylenedioxythiophene):polystyrene sulfonate (PEDOT:PSS). HTL consists of Poly((9,9-dioctylfluorene-2,7-diyl)-co-(4,40-N-(4-s-butylphenyl) diphenylamine) (TFB). In turn, the NRs are composed of a CdSe/ZnS core/shell structure, which is aligned along the z-axis. The ETL is composed of ZnO, while the cathode material is aluminum. Notably, the device architecture and scattering layer design together ensure that the photons generated in the EML are efficiently extracted along the z-axis, as reflected in the schematic and band diagrams. Importantly, the vertically aligned nanorods emit light that is predominantly polarized in the x/y (in-plane) directions, resulting in weak direct emission along the z-axis. To address this, our architecture includes a scattering layer above the emission region to redirect in-plane emission through the transparent ITO. This approach is purely optical and does not impact the electronic structure, charge transport, or quantum mechanical processes captured by our coupled drift-diffusion–Schrödinger–Poisson model.

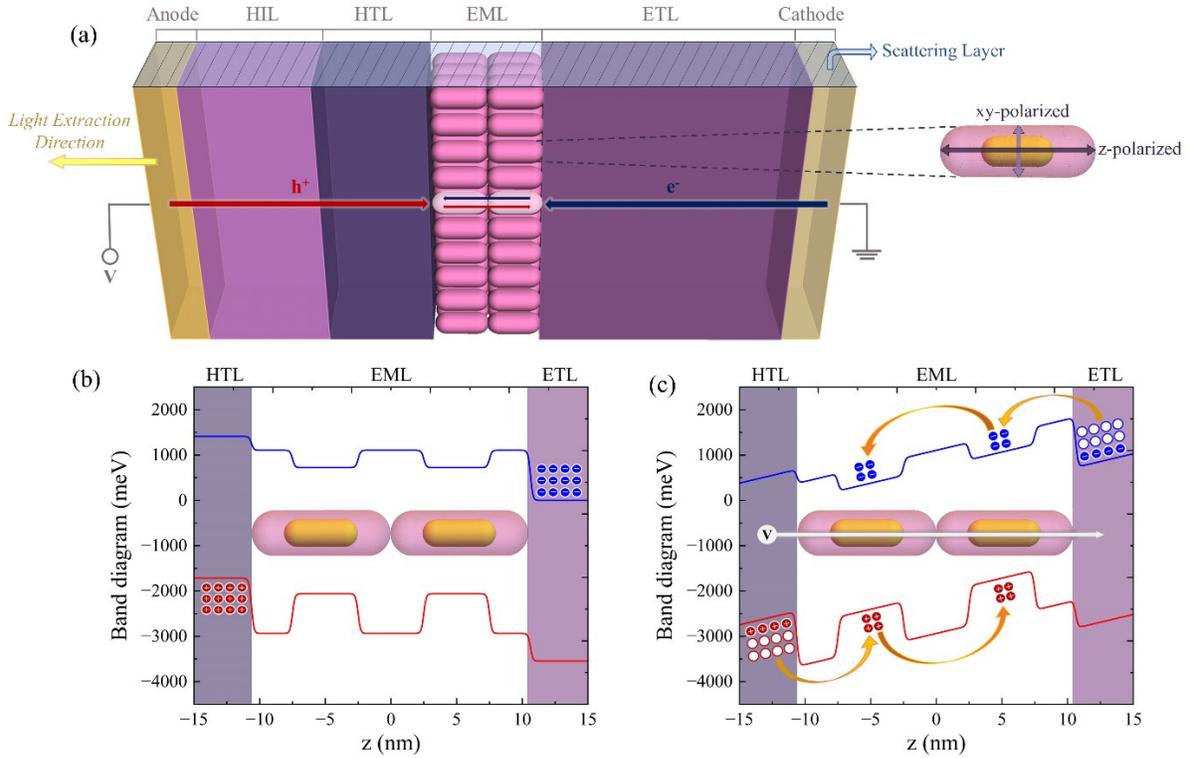

**Figure 1.** (a) Schematic of the NR-LED device architecture, showing each functional layer: HIL (PEDOT:PSS), HTL (TFB), EML (CdSe/ZnS nanorods), and ETL (ZnO), sandwiched between a transparent ITO anode and Al cathode. The device is surrounded by a scattering layer (on all faces except the ITO side) to enable efficient light extraction predominantly along the z-axis (normal to the substrate), as indicated by the yellow arrow. The inset illustrates emission polarization: intrinsic emission from vertically aligned nanorods is primarily in-plane (xy-polarized), but the scattering layer redirects this emission along z. (b) Band diagram of the full device under an applied bias equal to the built-in potential (i.e., the difference in electrode work functions), corresponding to the flat-band condition. (c) Band diagram under forward bias exceeding the built-in potential, illustrating electron and hole injection, transport, and radiative recombination within the EML.

Figures 1(b) and 1(c) display the corresponding band diagrams of the device in the absence and presence of a forward bias, respectively. In the flat-band scenario of Figure 1(b), which occurs at a bias equal to the built-in potential $\phi_{bi}$ (i.e., the work function difference of the electrodes), electrons and holes remain localized in the ETL and HTL, respectively, with minimal carrier injection into the EML. Upon applying a forward bias exceeding the built-in potential (Figure 1(c)), the band structure is tilted, facilitating efficient injection and transport of electrons and holes into the CdSe/ZnS nanorod layer. The resulting accumulation of carriers in the EML initiates radiative recombination, leading to photon emission. In our CdSe/ZnS rod-in-rod nanorod system, the adjacent CdSe cores are separated by two neighboring ZnS shells, each 2.5 nm thick, making the total barrier width 5 nm. For such a system, direct quantum tunneling through the shell barrier is recognized as the dominant charge transport mechanism, both theoretically and experimentally[49,50]. Thermionic emission and trap-assisted hopping are strongly suppressed for well-passivated nanorods at this barrier thickness, and studies have demonstrated that charge transfer in colloidal quantum dot and nanorod solids with nanometer-scale barriers is overwhelmingly determined by quantum tunneling [50]. In addition, state-of-the-art device models for QD-LEDs require explicit inclusion of interparticle tunneling to accurately reproduce experimental transport and emission characteristics [51]. Similar to the findings reported in the article [51], where it was demonstrated that the main transport mechanism between neighboring quantum dots is direct tunneling, in our case, the primary charge transport pathway between neighboring nanorods is also quantum tunneling.

This comprehensive configuration model incorporates the essential physics of charge transport, tunneling, drift-diffusion, recombination, and photon emission processes in NR-LEDs, thus providing a robust framework for analyzing and optimizing device performance.

## 2.2. Charge transport model

Charge transport within the device is categorized into three distinct mechanisms: drift-diffusion currents in the HIL/HTL and ETL, tunneling currents between adjacent NR layers, and injection currents at the interfaces between NR layers and charge transport layers. Drift-diffusion currents describe the movement of electrons and holes under the influence of an electric field (drift) and their random thermal motion (diffusion). Tunneling currents encompass electron and hole tunneling processes between the ETL and NR layers, as well as between the HTL and EML layers. Additionally, they include tunneling phenomena occurring between adjacent NRs.

The calculation of these current densities begins with solving the single-particle Schrödinger equation [52,53]:

$$\left[-\frac{\hbar^2}{2m_{e(h)}^*}\nabla^2 + V_{e(h)}(r,z) + q\big(\phi_{e(h)}(z) + \phi_{ext}(z)\big)\right]\Psi^{e(h)}(r,\theta,z) = E^{e(h)}\Psi^{e(h)}(r,\theta,z) \quad (1)$$

where, $m_{e(h)}^*$ represents the effective mass of the electron or hole, $V_{e(h)}(r,z)$ denotes the confinement potential energy for the electron or hole, $\phi_{e(h)}(z)$ is the electrostatic potential for the electron or hole obtained by the Poisson equation, $\phi_{ext}(z)$ is the external electrostatic potential of the system, $\Psi^{e(h)}(r,\theta,z)$ is the wave function of an electron or hole, and $E^{e(h)}$ is the energy. We can write $\Psi^{e(h)}(r,\theta,z) = \frac{e^{il\theta}}{\sqrt{2\pi}}R_n^{e(h)}(r)\psi_{nl}^{e(h)}(z)$, and solve the radial and axial eigenproblems separately. For the axial equations we will have:

$$\left[-\frac{\hbar^2}{2m_{e(h)}^*}\frac{d^2}{dz^2} + q\big(\phi_{e(h)}(z) + \phi_{ext}(z)\big)\right]\psi_{n\ell}^{e(h)}(z) = E_{n\ell}^{e(h)}\psi_{n\ell}^{e(h)}(z) \quad (2)$$

Using the obtained wave functions and energy levels, the local electron and hole densities can be calculated as follows [52,53]:

$$n(z) = 2\sum_n |\psi_{nl}^e(z)|^2 \, F\!\left(\frac{E_n^e - E_F^e(z)}{k_B T}\right)$$
$$p(z) = 2\sum_n |\psi_{nl}^h(z)|^2 \, F\!\left(\frac{E_F^h(z) - E_n^h}{k_B T}\right) \quad (3)$$

where, $n(z)$ and $p(z)$ represent the local charge densities of electrons and holes, respectively, as a function of position $z$, $F$ is the Fermi-Dirac distribution which describes the occupation probability of the system, $E_F^{e/h}(z)$ are the local quasi Fermi energy levels, $k_B$ is the Boltzmann constant, $T$ is the absolute temperature.

After obtaining the charge densities, the Poisson equation is solved to determine the electrostatic potential [54,55]:

$$\nabla \cdot [\varepsilon(z)\nabla\phi(z)] = -\rho(z) \quad (4)$$

where, $\phi(z)$ represents the electrostatic potential as a function of position $z$, $\rho(z)$ is the charge density, calculated as $\rho(z) = q \cdot \big(p(z) - n(z) + N_d^+ - N_a^-\big)$, $q$ is the elementary charge, $N_d^+$ and $N_a^-$ are the donor and acceptor doping densities of semiconductor layers, and $\varepsilon(z)$ is the permittivity of the material as a function of position $z$.

**Drift-Diffusion current:**

After solving the Schrödinger-Poisson system and obtaining the required quantities, such as carrier densities and the electrostatic potential, we can proceed to calculate the drift and diffusion currents in the system [51,56,57]. The Appendix A provides a detailed explanation of the algorithm, including the iterative process and all implemented steps. The drift current is generated by the motion of charge carriers (electrons and holes) under the influence of an electric field.

$$J_n^{\text{drift}}(z) = q \cdot n(z) \cdot \mu_n \cdot E(z)$$
$$J_p^{\text{drift}}(z) = q \cdot p(z) \cdot \mu_p \cdot E(z) \tag{5}$$

where, $E(z)$ represents the electric field as a function of position, calculated as $E(z) = -\dfrac{d\phi(z)}{dz}$, $\mu_n$ and $\mu_p$ are the electron and hole mobilities, respectively, which depend on the material properties and temperature.

The diffusion current arises due to the gradient in carrier concentration, driving carriers to move from regions of high concentration to regions of low concentration. This behavior is described by Fick's law as follows:

$$J_n^{\text{diff}}(z) = q \cdot D_n \cdot \frac{dn(z)}{dz}$$
$$J_p^{\text{diff}}(z) = -q \cdot D_p \cdot \frac{dp(z)}{dz} \tag{6}$$

where, $D_n$ and $D_p$ are electron and hole diffusion coefficients, related to the mobility by $D_{n,p} = \dfrac{k_B T}{q} \cdot \mu_{n,p}$.

The total drift-diffusion current is expressed as:

$$J_n(z) = J_n^{\text{drift}}(z) + J_n^{\text{diff}}(z)$$
$$J_p(z) = J_p^{\text{drift}}(z) + J_p^{\text{diff}}(z) \tag{7}$$

By substituting the respective expressions for drift and diffusion currents, the resulting equation can be obtained as follows:

$$J_n^D(z) = q \cdot n(z) \cdot \mu_n \cdot E(z) + q \cdot D_n \cdot \frac{dn(z)}{dz}$$
$$J_p^D(z) = q \cdot p(z) \cdot \mu_p \cdot E(z) - q \cdot D_p \cdot \frac{dp(z)}{dz} \tag{8}$$

**Tunneling current:**

In NR systems, charge transport between individual NRs does not occur through conventional band transport mechanisms. In the case of core-shell structured NRs, the shell serves as a narrow potential barrier, while the NR core functions as a potential well. This core-shell configuration limits the free movement of charge carriers and makes quantum tunneling the dominant mechanism for charge transfer between neighboring NRs. Quantum tunneling, governed by quantum mechanical principles, enables charge carriers to pass the potential barriers of the shell without requiring thermal activation. Instead, the tunneling process is facilitated by the finite width of the shell barrier and the quantum confinement effects of the NR core. The tunneling current density between two adjacent NRs can be effectively described by the following equation [58,59]:

$$J_n^T = q v_n d_{NR} (n_{(m+1)} - n_m)$$
$$J_p^T = -q v_p d_{NR} (p_{(m+1)} - p_m) \tag{9}$$

where, $v_n$ and $v_p$ represent the tunneling frequencies respectively, $d_{NR}$ denotes the NR diameter, while $n_{(m+1)}$, $n_m$, $p_{(m+1)}$, $p_m$ correspond to the electron and hole densities at the NR layers indexed by $m$ and $m+1$. The tunneling frequency for a charge carrier oscillating between two adjacent NRs can be expressed using the following equation:

$$v = \frac{v_{th}}{2 d_{NR}} T_{bs} \tag{10}$$

where, $v_{th}$ is a thermal velocity of a given carrier in NR core, that is calculated as follows:

$$v_{th} = \left( \frac{2 k_B T}{m_{e(h)}^*} \right)^{1/2} \tag{11}$$

The $2k_B T$ in Eq. 10 comes from the two free translational directions—axial (which includes tunneling) and azimuthal. Tunneling is therefore treated as part of the axial motion [51].

The tunneling probability $T_{bs}$ through the NR shells between two neighboring NRs, when a carrier crosses a tunneling barrier formed by a NR shell of thickness $t_s$, is described by the following equation:

$$T_{bs} = \begin{cases} 1, & \Delta E^S < 0 \\ \exp\left(-\frac{2\pi}{h} t_s \sqrt{8 m_s^* \Delta E^S}\right), & \Delta E^S \geq 0 \end{cases} \quad (12)$$

The energy barrier height, $\Delta E^S$, is defined as the difference between the energy level of the shell ($E^S$) and the energy level of the carrier ($E$), such that $\Delta E^S = E^S - E$.

**Injection current:**

The charge injection current density under forward bias conditions, occurring between the NR layer and the charge transport layers, can be determined using the following equation [51]:

$$\begin{aligned} J_n^I &= q r_{NR} (\pi r_{NR}^2) \sigma_n T_{bn} \mu_{n0}^{NR} \left(\frac{F_n^{3/2}}{F_0^{1/2}}\right) n_E (N_{NR} - n_{Q2}) \\ J_p^I &= q r_{NR} (\pi r_{NR}^2) \sigma_p T_{bp} \mu_{p0}^{NR} \left(\frac{F_p^{3/2}}{F_0^{1/2}}\right) p_H (N_{NR} - p_{Q1}) \end{aligned} \quad (13)$$

where, $r_{NR}$ denotes the radius of a NR, the term $\pi r_{NR}^2$ corresponds to the cross-sectional area of the NR, while $\sigma_n$ and $\sigma_p$ are the relative capture cross-sections for electrons and holes, respectively. The parameters $\mu_{n0}^{NR}$ and $\mu_{p0}^{NR}$ indicate the mobilities of electrons and holes within the NR layer under a reference electric field $F_0$, as defined by the Poole–Frenkel law. The $F_n$ and $F_p$ are electric fields at the interfaces of the ETL and the NR layer, and HTL and the NR layer. The hole and electron densities at the ETL and HTL surfaces adjacent to the NR layer, are represented as $n_E$ and $p_H$ respectively, which are dependent on the applied voltage and serve as sources of holes and electrons for the neighboring NR layer. The parameters $n_{Q2}$ and $p_{Q1}$ denote the electron and hole densities at the centers of the outermost NR layers facing the ETL and HTL, respectively. The density of NRs ($N_{NR}$) is determined as the reciprocal of the volume of a single NR particle.

To simplify the analysis of charge injection imbalance on charge distribution in NR layers, the charge injection mobility parameters $\alpha_n = \sigma_n T_{bn} \mu_{n0}^{NR}$ and $\alpha_p = \sigma_p T_{bp} \mu_{p0}^{NR}$ are used. By incorporating these expressions into the Eq. 12, the injection current density equations take the following form:

$$J_n^I = qr_{NR}(\pi r_{NR}^2)\alpha_n \left(\frac{F_n^{3/2}}{F_0^{1/2}}\right) n_E (N_{NR} - n_{Q2})$$
$$J_p^I = qr_{NR}(\pi r_{NR}^2)\alpha_p \left(\frac{F_p^{3/2}}{F_0^{1/2}}\right) p_H (N_{NR} - p_{Q1})$$
(14)

**Total current:**

The total current across the NR-LED device can be expressed as the sum of contributions from drift-diffusion, tunneling, and injection currents:

$$J_n = J_n^D + J_n^T + J_n^I$$
$$J_p = J_p^D + J_p^T + J_p^I$$
(15)

### 3. Results

In this section, we will first delve into the alignment of the conduction band in our system under different biases, as illustrated in Figure 2. We won't show the same dynamics for the valence band, as the behavior is similar for both bands. The lower section of the figure depicts the structure of the conduction band in the absence of an external voltage, while the uppermost section shows the band structure under an applied voltage of 0.68 V. The conduction band offset between the HTL and the NR is also examined to evaluate the impact of charge injection mismatch resulting from the disparity in conduction band offsets. For the numerical calculations, we utilized material parameters presented in Table 1, which is provided in Appendix B.

Electron probability densities for both ground and first excited states under various external voltages are shown in Figure 2 (a) and (b). Selected voltages illustrate the sequential localization shift of electrons from the ETL region, through the EML region, and ultimately to the HTL region. Without applied voltage, the electron probability distributions exhibit symmetry, but under applied bias, distributions become notably asymmetric due to structural and potential landscape asymmetries.

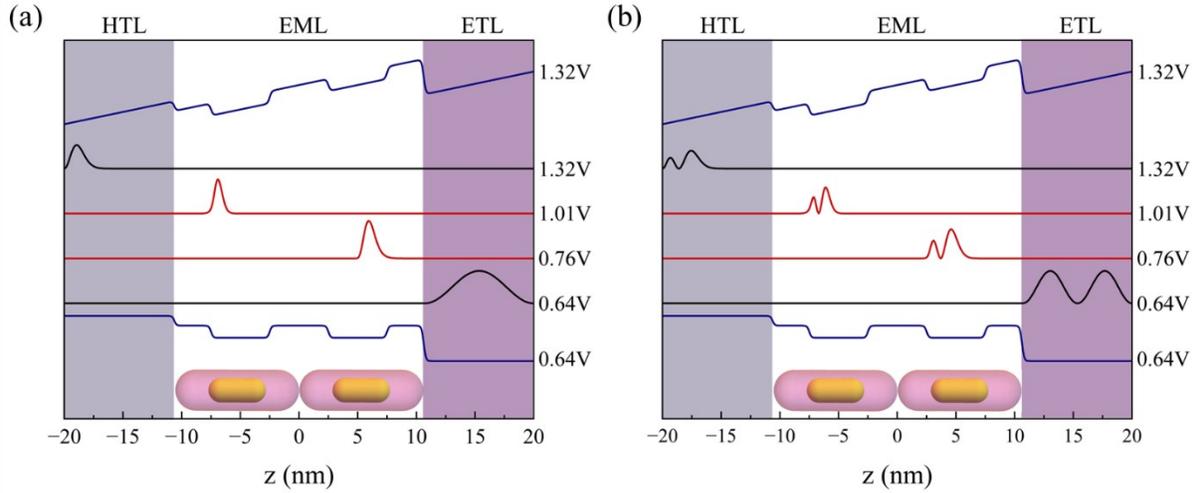

**Figure 2.** Band structure of the system and electron probability distributions for the ground state (a) and the first excited state (b) at different values of external voltage.

The dynamics of electron localization and tunneling are further detailed through three-dimensional static probability distributions depicted in Figure 3. Initially, electrons are absent from both nanorods (NRs). Upon applying a bias voltage, electrons localize first within the right NR (ETL), presenting an asymmetric distribution shaped by the spatial arrangement and confinement potentials. Increasing voltage facilitates quantum tunneling across the potential barrier to the left NR (HTL). Tunneling efficiency and timescale strongly depend on NR spacing and confinement potential, indicating the tunability of these parameters for electronic applications.

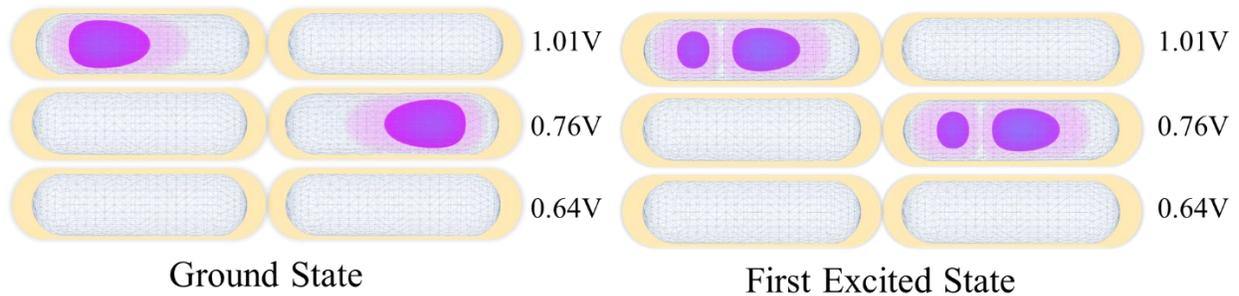

**Figure 3.** Probability distributions of a single electron for the ground and first excited states under three different applied voltage values.

Once electrons reach the left NR, asymmetry in the probability distribution persists, underscoring how intrinsic structural features continually influence electron dynamics. Moreover, the applied electric field significantly modifies electron localization and induces a Stark shift in the electron energy spectrum.

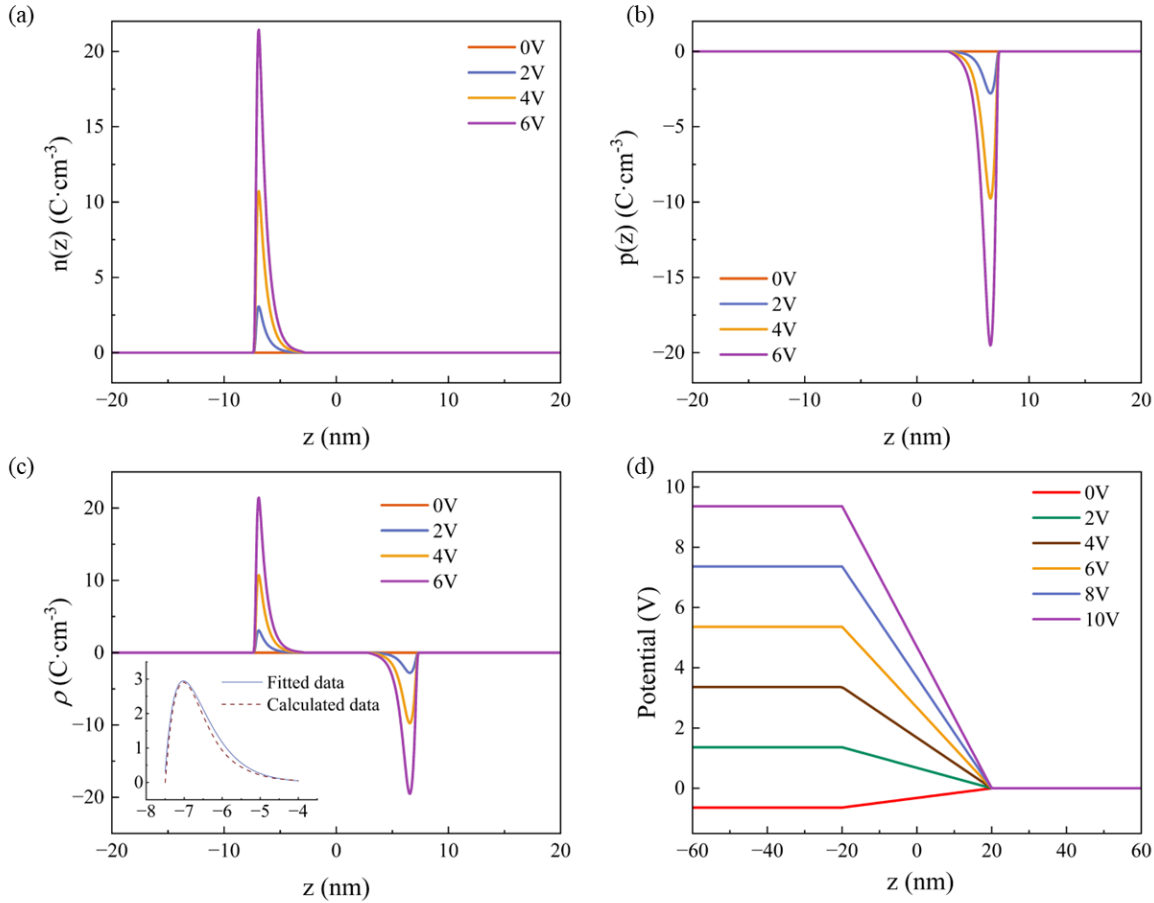

**Figure 4.** Internal distributions of electron density (a), hole density (b), charge density (c) and electrostatic potential (d) within the system under various voltage configurations.

Building on this, the behavior of the carrier probability densities under an applied voltage is reflected in various physical quantities such as electron density, hole density, charge density charge density, and electrostatic potential, as illustrated in Figure 4. Electron, hole and charge density distributions across the quantized NR region (EML) were calculated under various applied voltages and are presented in Figure 4(a-c). It is demonstrated that positive and negative charges accumulate on the surfaces of the HTL and ETL adjacent to the NR layer, thereby altering the electrostatic potential and generating internal electric fields within the NR layer. The charge distribution near the HTL/EML or EML/ETL boundaries, corresponding to electrons and holes respectively, exhibits an asymmetric form. This distribution can be accurately modeled using the well-known asymmetric Erlang distribution function, expressed as $f(z;k,\lambda) = \dfrac{\lambda^k z^{k-1} e^{-\lambda z}}{(k-1)!}$, where $k > 0$ is an integer shape parameter, and $\lambda > 0$ is a real-valued rate parameter. For each applied bias voltage, the parameters $k$ and $\lambda$ can be fitted. The resulting fit for the 2V case is shown in the inset of Figure 4(c). The electrostatic potentials were computed for bias voltages ranging from 0 V to 10 V and are presented in Figure 4(d). The figure reveals that the potential decreases linearly only within the EML region, resulting in a concentrated electric field in the NR layers. The potential drops linearly within the emissive layer and facilitates charge injection by reducing energy barriers at interfaces, enhancing the efficiency of carrier transfer between layers.

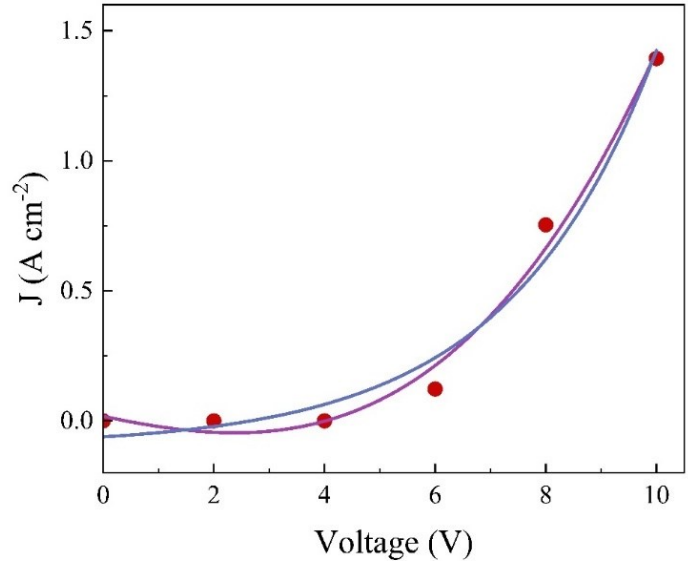

**Figure 5.** Current-voltage (I-V) characteristics of the double NR system.

Finally, by calculating the current densities of all components, we obtain the current-voltage (I-V) characteristics of the system under consideration. The dependence of the current density on the applied voltage is presented in Figure 5. The data points represent the calculated values, while the fitted curves illustrate the corresponding analytical approximations. The first curve (blue) is fitted using an exponential function of the form $J \approx \alpha + \beta e^{\gamma u}$, whereas the second curve (purple) is approximated by a polynomial function $J \approx \alpha + \beta u + \gamma u^2 + \delta u^3$. The results demonstrate a nonlinear dependence, where the current density remains nearly constant at low bias voltages. However, beyond approximately 4V, a noticeable increase is observed, which can be well described by an exponential function, indicating a transition to a regime dominated by charge injection and transport mechanisms. The apparent turn-on at 4 V is due to carrier injection barriers: at biases below 4 V, electrons and holes cannot overcome the energy offsets at the ETL/NR and HTL/NR interfaces imposed by the band alignment. Once the applied bias reaches 4 V, the electrostatic potential drop compensates these interfacial barriers, enabling efficient carrier injection into the NR layers.

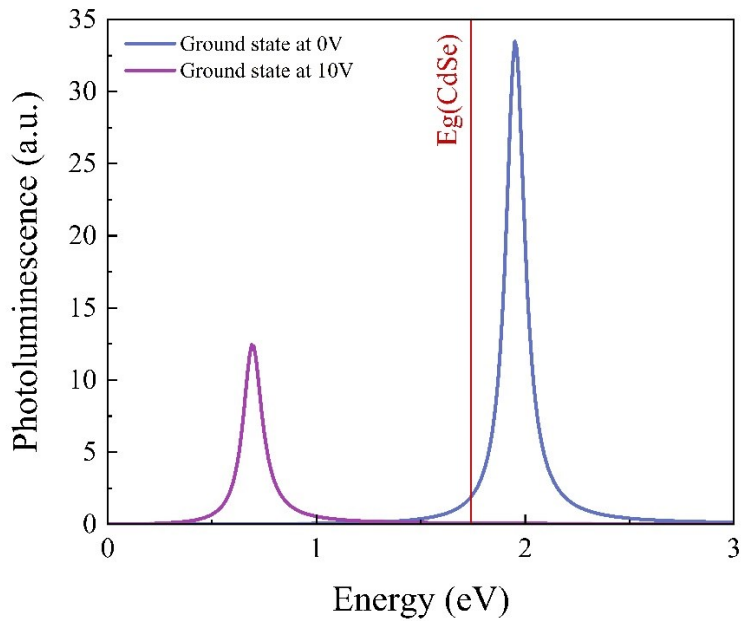

**Figure 6.** Dependence of the photoluminescence spectra on the presence and absence of an external bias voltage in the NR-LED system.

Figure 6 illustrates the dependence of the photoluminescence (PL) spectra on the presence and absence of an external bias voltage in the NR-LED system. The detailed methodology for calculating the PL spectra is provided in Appendix C. In our analysis, we consider all possible transitions between the first six electron and hole states. The corresponding oscillator strengths for each transition are calculated and incorporated into the total spectra. For reference, the band gap of CdSe is also indicated in the figure. The total PL spectrum is obtained by summing the contributions from all accounted transitions, after considering the non-equilibrium population of each state all the transitions are eliminated except the fundamental interband transition between the ground electron and ground hole states. Notably, in the absence of an external voltage, this transition appears at a higher energy due to the intrinsic quantum confinement effects. However, upon the application of a 10V bias, this transition experiences a redshift, moving to lower energy values, ultimately falling below the classical band gap energy of CdSe.

A comparative analysis of the PL spectra reveals two primary effects induced by the applied voltage: the entire spectrum shifts towards lower energy values, indicating a decrease in transition energies, and the overall intensity of the PL spectra decreases when an external voltage is applied, suggesting modifications in the recombination dynamics. These observations demonstrate that an external bias voltage serves as an effective tuning mechanism for controlling both the emission energy and intensity of the NR-LED system.

## 4. Conclusion

In this work, we developed a comprehensive theoretical model to describe charge transport mechanisms in NR-based LEDs. By incorporating drift-diffusion, tunneling, and injection currents, we provided a detailed analysis of carrier dynamics within the multilayer NR-LED architecture. Our findings demonstrate that the interplay between charge injection, tunneling, and recombination significantly influences the device performance, highlighting the importance of understanding carrier transport mechanisms at the nanoscale.

The numerical simulations reveal that applied external voltage strongly affects the band structure, inducing charge redistribution across the NR layers. This leads to stepwise localization of electrons and holes, which is primarily governed by quantum tunneling processes. The study also shows that charge transport is highly asymmetric due to differences in conduction band offsets between adjacent layers, which directly impact carrier mobility and recombination rates.

The current-voltage characteristics obtained from our model confirm a nonlinear dependence of current density on the applied voltage. At lower bias voltages, charge transport is limited by injection barriers, while at higher voltages, an exponential increase in current density is observed due to enhanced tunneling and drift-diffusion transport. This behavior suggests that optimizing the thickness and material composition of the transport and emission layers can lead to improved efficiency and performance in NR-LEDs.

In conclusion, the analysis of the PL spectra in the NR-LED system reveals significant changes due to the application of an external bias voltage. The observed redshift in the fundamental interband transition, moving below the classical CdSe band gap, indicates the influence of the external voltage on quantum confinement effects. Moreover, the comparative analysis of the PL spectra demonstrates two key effects: a shift of the entire spectrum towards lower energy values and a reduction in overall intensity. These changes suggest that the applied voltage modifies the transition energies within the system.

Overall, our study provides a robust theoretical framework for understanding and optimizing charge transport in NR-based optoelectronic devices. The insights gained from this work can be instrumental in guiding the design of advanced nanostructured LEDs with enhanced efficiency and tailored optical properties. Future research directions may include the incorporation of many-body effects, excitonic interactions, and temperature-dependent transport mechanisms to further refine the predictive capabilities of our model. Additionally, experimental validation of our theoretical predictions would be an essential next step toward realizing high-performance NR-LEDs for practical applications in optoelectronics and display technologies.

**Acknowledgement**

This work was supported by the Science Committee of RA (Research project № 24LCG-1C005). A.G. Melkonyan expresses her sincere gratitude to the Quanta "Quantum and Mesoscopic Physics" Master's Degree Program for the scholarship and support provided throughout her studies. This research, which forms the basis of her Master's thesis, has been significantly enriched by the opportunities and resources offered by the program, contributing profoundly to his academic and professional development.

**Appendix A**

In this section, we present the algorithm used for calculating wave functions and energies in the NR-LED system. The process involves an iterative self-consistent solution of the Schrödinger-Poisson equation for current density calculations.

The procedure follows these steps: Initially, the electron and hole energy levels and wave functions are computed in the potential defined solely by the band-offsets, neglecting the self-consistent potential. Using these calculated wave functions and energy levels, we then determine the local charge densities of electrons and holes. These charge distributions allow us to solve the Poisson equation and obtain the self-consistent electrostatic potential that arises in the system.

In the next iteration, the Schrödinger equation is solved again, this time incorporating the self-consistent potential obtained from the previous step. The updated wave functions and energy levels are then recalculated for both electrons and holes. This iterative procedure continues, refining the self-consistent potential at each step.

After a sufficient number of iterations, the self-consistent potential converges to a stable solution, meaning it no longer changes significantly between iterations. At this stage, we proceed to the final step, where all relevant current densities, including their various contributing components, are computed along with the total current density of the system.

In our computational approach, the simulation domain extends beyond the NR boundary to accommodate the spatial extent of electron and hole wave functions, allowing them to penetrate and gradually decay into the surrounding free space. This is particularly important for accurately capturing the quantum confinement effects and the influence of the external potential.

To ensure proper boundary treatment, we impose von Neumann boundary conditions at the internal interface between the NRs core and shell regions, as well as at the outer boundary of the NR. This choice is dictated by the Ben-Daniel-Duke boundary conditions, which ensure the continuity of both the wave function and its derivative in systems with spatially varying effective mass. The Ben-Daniel-Duke formalism is particularly relevant in core/shell heterostructures, where the mass discontinuity can significantly impact the energy eigenstates and tunneling properties of charge carriers.

# Iterative Self-consistent Calculation Of The Schrödinger–Poisson Equation For Current Density Calculation

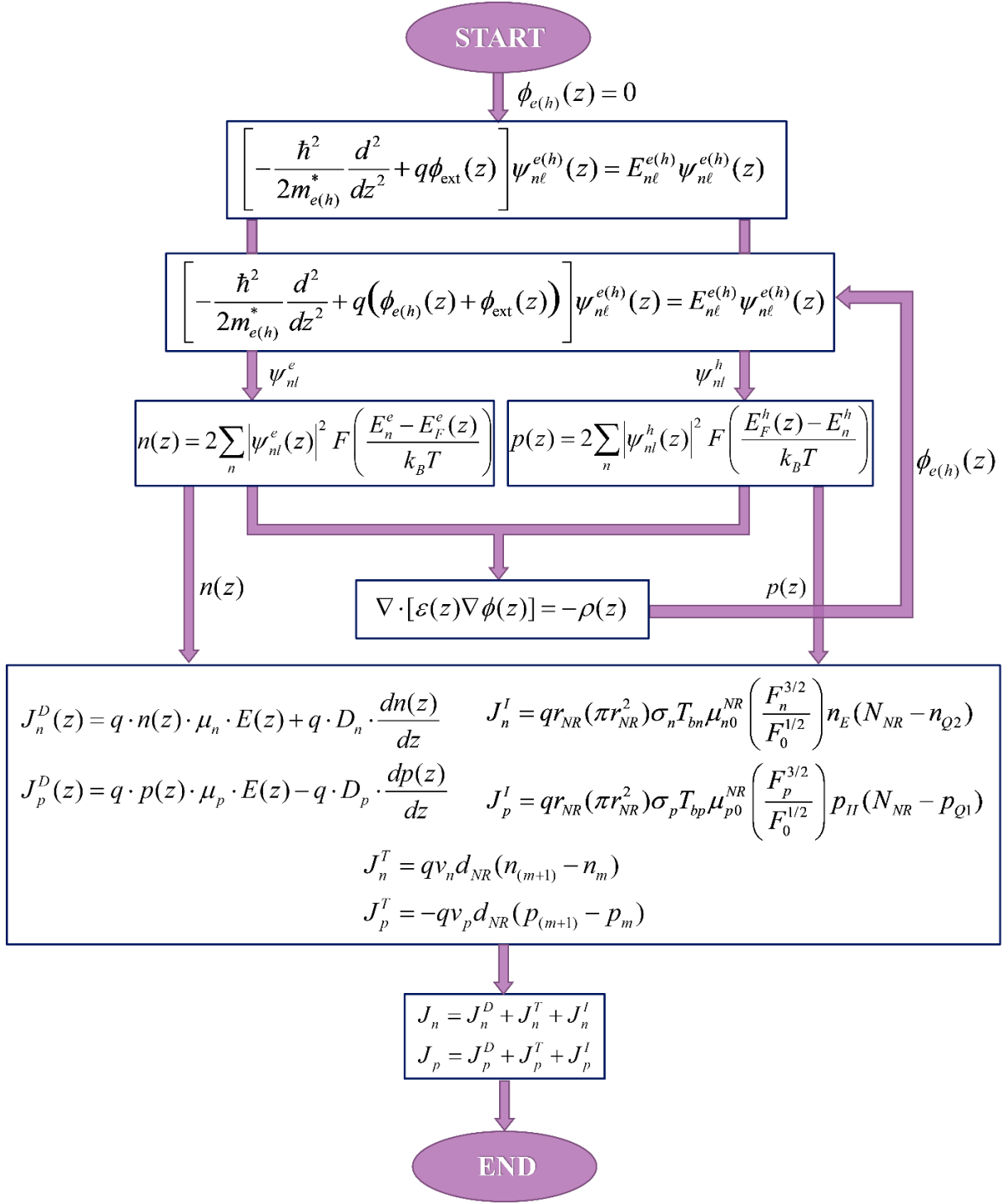

**Figure A.** Schematics of the algorithm used to calculate the total current density of the NR-LED system.

At the outermost edge of the computational domain, we apply Dirichlet boundary conditions by setting the wave function to zero. This choice effectively simulates the presence of an infinite potential barrier, preventing unphysical reflections and ensuring that the wave function smoothly decays without artificial constraints. The selected domain size is sufficient to ensure that the wave function vanishes naturally within the given computational region, avoiding spurious effects that could arise from truncation artifacts.

This boundary configuration enables an accurate description of the quantum states while maintaining numerical stability and minimizing errors associated with the finite computational domain. Additionally, it provides a realistic approximation of carrier confinement and tunneling processes, which are crucial for studying optical transitions, excitonic effects, and transport phenomena in NR-LED structures.

## Appendix B

**Table1.** Materials parameters used in the calculations [51].

| | Nanorod Parameters | | Materials and Device Parameters | | | | |
|---|---|---|---|---|---|---|---|
| | Core | Shell | Anode | HIL | HTL | ETL | Cathode |
| Materials | CdSe | ZnS | ITO | PEDOT:PSS | TFB | ZnO | Al |
| Diameter (nm) | 5 | 2.5 | | 20 | 20 | 40 | |
| Electron mobility, $\mu_n$ (cm$^2$V$^{-1}$s$^{-1}$) | 2×10$^{-6}$ | | | | | 2×10$^{-3}$ | |
| Hole mobility, $\mu_p$ (cm$^2$V$^{-1}$s$^{-1}$) | 1×10$^{-6}$ | | | 0.322×10$^{-6}$ | 2×10$^{-3}$ | | |
| Electron effective mass, $m_n^*/m_0$ | 0.13 | 0.19 | | 1 | 1 | 0.24 | |
| Hole effective mass, $m_p^*/m_0$ | 0.45 | 0.6 | | 1 | 1 | 0.59 | |

| Charge injection mobility, $\alpha_n = \alpha_p$ (cm$^2$V$^{-1}$s$^{-1}$) | 3×10$^{-9}$ | | | | | | |
|---|---|---|---|---|---|---|---|
| Doping type | | | | p-type | p-type | n-type | |
| Doping concentrations, $N_a$, $N_d$ (cm$^{-3}$) | | | | 2.81×10$^{19}$ | 1×10$^{17}$ | 1×10$^{17}$ | |
| Work-functions, $\phi_{wf}$ [eV] | | | 4.7 | | | | 4.06 |

## Appendix C

The calculation of PL spectra for our NR-LED system is done by using Roosbroeck-Shockley relation [60,61]:

$$R(\hbar\omega) = R_0 \hbar\omega K(\hbar\omega) \frac{f_c(1-f_v)}{f_v - f_c} \tag{C1}$$

where $R_0$ is proportional to the square modulus of the dipole moment matrix element taken over Bloch functions, $f_c$ and $1-f_v$ represent the occupancy probability of the conduction band, and the empty state probability of the valance band, respectively. $K(\hbar\omega)$ is the absorption coefficient which can be written as:

$$K(\hbar\omega) = \frac{A_0}{V} \sum_{n,n'} \left| \int_\Omega \psi_e(\vec{r}_e) \psi_h(\vec{r}_h) d\vec{r}_e d\vec{r}_h \right|^2 \delta\left(\hbar\omega - \left(E_g + E_e + E_h\right)\right) \tag{C2}$$

where $n$ and $n'$ denote sets of quantum numbers, coefficient $A_0$ is taken on the Bloch functions and proportional to the square of the dipole matrix element [62,63].

To handle the energy-conserving delta function we approximate it with a Lorentzian function, which can be expressed as:

$$\delta\left(\hbar\omega - \left(E_g + E_e + E_h\right)\right) = \frac{\hbar\Gamma}{\pi\left(\left(\hbar\omega - \left(E_g + E_e + E_h\right)\right)^2 + (\hbar\Gamma)^2\right)} \tag{C3}$$

where $\Gamma$ is the line width of the exciton [64].

# References

1. Kagan, C.R., Lifshitz, E., Sargent, E.H. and Talapin, D.V., 2016. Building devices from colloidal quantum dots. *Science*, *353*(6302), p.aac5523.
2. Zhou, Y., Zhao, H., Ma, D. and Rosei, F., 2018. Harnessing the properties of colloidal quantum dots in luminescent solar concentrators. *Chemical Society Reviews*, *47*(15), pp.5866-5890.
3. Hens, Z. and Moreels, I., 2012. Light absorption by colloidal semiconductor quantum dots. *Journal of Materials Chemistry*, *22*(21), pp.10406-10415.
4. Qin, H., Meng, R., Wang, N. and Peng, X., 2017. Photoluminescence Intermittency and Photo-Bleaching of Single Colloidal Quantum Dot. *Advanced Materials*, *29*(14), p.1606923.
5. Shirasaki, Y., Supran, G.J., Bawendi, M.G. and Bulović, V., 2013. Emergence of colloidal quantum-dot light-emitting technologies. *Nature photonics*, *7*(1), pp.13-23.
6. Kim, J., Roh, J., Park, M. and Lee, C., 2024. Recent advances and challenges of colloidal quantum dot light-emitting diodes for display applications. *Advanced Materials*, *36*(20), p.2212220.
7. Kagan, C.R., Bassett, L.C., Murray, C.B. and Thompson, S.M., 2020. Colloidal quantum dots as platforms for quantum information science. *Chemical reviews*, *121*(5), pp.3186-3233.
8. Litvin, A.P., Martynenko, I.V., Purcell-Milton, F., Baranov, A.V., Fedorov, A.V. and Gun'Ko, Y.K., 2017. Colloidal quantum dots for optoelectronics. *Journal of Materials Chemistry A*, *5*(26), pp.13252-13275.
9. Liu, M., Yazdani, N., Yarema, M., Jansen, M., Wood, V. and Sargent, E.H., 2021. Colloidal quantum dot electronics. *Nature Electronics*, *4*(8), pp.548-558.
10. Luo, J., Ma, L., He, T., Ng, C.F., Wang, S., Sun, H. and Fan, H.J., 2012. TiO2/(CdS, CdSe, CdSeS) nanorod heterostructures and photoelectrochemical properties. *The Journal of Physical Chemistry C*, *116*(22), pp.11956-11963.
11. Zhang, C., Chen, J., Kong, L., Wang, L., Wang, S., Chen, W., Mao, R., Turyanska, L., Jia, G. and Yang, X., 2021. Core/shell metal halide perovskite nanocrystals for optoelectronic applications. *Advanced Functional Materials*, *31*(19), p.2100438.

12. Sitt, A., Hadar, I. and Banin, U., 2013. Band-gap engineering, optoelectronic properties and applications of colloidal heterostructured semiconductor nanorods. *Nano Today*, *8*(5), pp.494-513.
13. Mehata, M.S. and Ratnesh, R.K., 2019. Luminescence properties and exciton dynamics of core–multi-shell semiconductor quantum dots leading to QLEDs. *Dalton Transactions*, *48*(22), pp.7619-7631.
14. Ji, B., Koley, S., Slobodkin, I., Remennik, S. and Banin, U., 2020. ZnSe/ZnS core/shell quantum dots with superior optical properties through thermodynamic shell growth. *Nano letters*, *20*(4), pp.2387-2395.
15. Zhang, C., Chen, J., Kong, L., Wang, L., Wang, S., Chen, W., Mao, R., Turyanska, L., Jia, G. and Yang, X., 2021. Core/shell metal halide perovskite nanocrystals for optoelectronic applications. *Advanced Functional Materials*, *31*(19), p.2100438.
16. Zhang, D., Dong, G., Cao, Y. and Zhang, Y., 2018. Ethanol gas sensing properties of lead sulfide quantum dots-decorated zinc oxide nanorods prepared by hydrothermal process combining with successive ionic-layer adsorption and reaction method. *Journal of colloid and interface science*, *528*, pp.184-191.
17. Pandi, D.V., Muthukumarasamy, N., Agilan, S. and Velauthapillai, D., 2018. CdSe quantum dots sensitized ZnO nanorods for solar cell application. *Materials Letters*, *223*, pp.227-230.
18. Gwo, S., Lu, Y.J., Lin, H.W., Kuo, C.T., Wu, C.L., Lu, M.Y. and Chen, L.J., 2017. Nitride semiconductor nanorod heterostructures for full-color and white-light applications. In *Semiconductors and Semimetals* (Vol. 96, pp. 341-384). Elsevier.
19. Yang, G., He, Y., Zhao, J., Chen, S. and Yuan, R., 2021. Ratiometric electrochemiluminescence biosensor based on Ir nanorods and CdS quantum dots for the detection of organophosphorus pesticides. *Sensors and Actuators B: Chemical*, *341*, p.130008.
20. Sarangi, S.N., Nozaki, S. and Sahu, S.N., 2015. ZnO nanorod-based non-enzymatic optical glucose biosensor. *Journal of biomedical nanotechnology*, *11*(6), pp.988-996.
21. Davis, N.J., Böhm, M.L., Tabachnyk, M., Wisnivesky-Rocca-Rivarola, F., Jellicoe, T.C., Ducati, C., Ehrler, B. and Greenham, N.C., 2015. Multiple-exciton generation in lead selenide nanorod solar cells with external quantum efficiencies exceeding 120%. *Nature communications*, *6*(1), p.8259.

22. Jeltsch, K.F., Schädel, M., Bonekamp, J.B., Niyamakom, P., Rauscher, F., Lademann, H.W., Dumsch, I., Allard, S., Scherf, U. and Meerholz, K., 2012. Efficiency enhanced hybrid solar cells using a blend of quantum dots and nanorods. *Advanced Functional Materials*, *22*(2), pp.397-404.
23. Bhatt, S., Shukla, R., Pathak, C. and Pandey, S.K., 2021. Evaluation of performance constraints and structural optimization of a core-shell ZnO nanorod based eco-friendly perovskite solar cell. *Solar Energy*, *215*, pp.473-481.
24. Park, Y.S., Roh, J., Diroll, B.T., Schaller, R.D. and Klimov, V.I., 2021. Colloidal quantum dot lasers. *Nature Reviews Materials*, *6*(5), pp.382-401.
25. Jung, H., Ahn, N. and Klimov, V.I., 2021. Prospects and challenges of colloidal quantum dot laser diodes. *Nature Photonics*, *15*(9), pp.643-655.
26. Minotto, A., Haigh, P.A., Łukasiewicz, Ł.G., Lunedei, E., Gryko, D.T., Darwazeh, I. and Cacialli, F., 2020. Visible light communication with efficient far-red/near-infrared polymer light-emitting diodes. *Light: Science & Applications*, *9*(1), p.70.
27. Cahyadi, W.A. and Chung, Y.H., 2018. Experimental demonstration of indoor uplink near-infrared LED camera communication. *Optics express*, *26*(15), pp.19657-19664.
28. Chen, Y., Wang, S. and Zhang, F., 2023. Near-infrared luminescence high-contrast in vivo biomedical imaging. *Nature Reviews Bioengineering*, *1*(1), pp.60-78.
29. Yu, H., Chen, J., Mi, R., Yang, J. and Liu, Y.G., 2021. Broadband near-infrared emission of K3ScF6: Cr3+ phosphors for night vision imaging system sources. *Chemical Engineering Journal*, *417*, p.129271.
30. Vasilopoulou, M., Fakharuddin, A., García de Arquer, F.P., Georgiadou, D.G., Kim, H., Mohd Yusoff, A.R.B., Gao, F., Nazeeruddin, M.K., Bolink, H.J. and Sargent, E.H., 2021. Advances in solution-processed near-infrared light-emitting diodes. *Nature Photonics*, *15*(9), pp.656-669.
31. Zhang, Y., Zhang, F., Wang, H., Wang, L., Wang, F., Lin, Q., Shen, H. and Li, L.S., 2019. High-efficiency CdSe/CdS nanorod–based red light–emitting diodes. Optics express, 27(6), pp.7935-7944.

32. Jiang, Y., Cho, S.Y. and Shim, M., 2018. Light-emitting diodes of colloidal quantum dots and nanorod heterostructures for future emissive displays. *Journal of Materials Chemistry C*, *6*(11), pp.2618-2634.
33. Kang, J.W., Kim, B.H., Song, H., Jo, Y.R., Hong, S.H., Jung, G.Y., Kim, B.J., Park, S.J. and Cho, C.H., 2018. Radial multi-quantum well ZnO nanorod arrays for nanoscale ultraviolet light-emitting diodes. *Nanoscale*, *10*(31), pp.14812-14818.
34. Srivastava, A.K., Zhang, W., Schneider, J., Halpert, J.E. and Rogach, A.L., 2019. Luminescent down-conversion semiconductor quantum dots and aligned quantum rods for liquid crystal displays. *Advanced Science*, *6*(22), p.1901345.
35. Lin, H.W., Lu, Y.J., Chen, H.Y., Lee, H.M. and Gwo, S., 2010. InGaN/GaN nanorod array white light-emitting diode. *Applied Physics Letters*, *97*(7).
36. Lu, J., Shi, Z., Wang, Y., Lin, Y., Zhu, Q., Tian, Z., Dai, J., Wang, S. and Xu, C., 2016. Plasmon-enhanced electrically light-emitting from ZnO nanorod arrays/p-GaN heterostructure devices. *Scientific Reports*, *6*(1), p.25645.
37. Park, H.K., Yoon, S.W., Eo, Y.J., Chung, W.W., Yoo, G.Y., Oh, J.H., Lee, K.N., Kim, W. and Do, Y.R., 2016. Horizontally assembled green InGaN nanorod LEDs: scalable polarized surface emitting LEDs using electric-field assisted assembly. *Scientific reports*, *6*(1), p.28312.
38. Kim, S., Lee, H., Jung, G.H., Kim, M., Kim, I., Han, M., Lee, S., Oh, S., Lim, J.H. and Kim, K.K., 2023. Self-array of one-dimensional GaN nanorods using the electric field on dielectrophoresis for the photonic emitters of display pixel. *Nanoscale Advances*, *5*(4), pp.1079-1085.
39. Wang, M., Ye, C.H., Zhang, Y., Hua, G.M., Wang, H.X., Kong, M.G. and Zhang, L.D., 2006. Synthesis of well-aligned ZnO nanorod arrays with high optical property via a low-temperature solution method. *Journal of Crystal Growth*, *291*(2), pp.334-339.
40. Choi, M.K., Yang, J., Hyeon, T. and Kim, D.H., 2018. Flexible quantum dot light-emitting diodes for next-generation displays. *npj Flexible Electronics*, *2*(1), p.10.
41. Srivastava, A.K., Zhang, W., Schneider, J., Halpert, J.E. and Rogach, A.L., 2019. Luminescent down-conversion semiconductor quantum dots and aligned quantum rods for liquid crystal displays. *Advanced Science*, *6*(22), p.1901345.

42. Ha, S.T., Fu, Y.H., Emani, N.K., Pan, Z., Bakker, R.M., Paniagua-Domínguez, R. and Kuznetsov, A.I., 2018. Directional lasing in resonant semiconductor nanoantenna arrays. *Nature nanotechnology*, *13*(11), pp.1042-1047.
43. Zhao, K., Pan, Z. and Zhong, X., 2016. Charge recombination control for high efficiency quantum dot sensitized solar cells. *The journal of physical chemistry letters*, *7*(3), pp.406-417.
44. Park, Y.S., Bae, W.K., Baker, T., Lim, J. and Klimov, V.I., 2015. Effect of Auger recombination on lasing in heterostructured quantum dots with engineered core/shell interfaces. *Nano Letters*, *15*(11), pp.7319-7328.
45. Reddeppa, M., Park, B.G., Majumder, S., Kim, Y.H., Oh, J.E., Kim, S.G., Kim, D. and Kim, M.D., 2020. Hydrogen passivation: a proficient strategy to enhance the optical and photoelectrochemical performance of InGaN/GaN single-quantum-well nanorods. *Nanotechnology*, *31*(47), p.475201.
46. Luo, N., Liao, G. and Xu, H.K., 2016. kp theory of freestanding narrow band gap semiconductor nanowires. *AIP Advances*, *6*(12).
47. Sengupta, P., Ryu, H., Lee, S., Tan, Y. and Klimeck, G., 2016. Numerical guidelines for setting up a kp simulator with applications to quantum dot heterostructures and topological insulators. *Journal of Computational Electronics*, *15*, pp.115-128.
48. Hönig, G., Callsen, G., Schliwa, A., Kalinowski, S., Kindel, C., Kako, S., Arakawa, Y., Bimberg, D. and Hoffmann, A., 2014. Manifestation of unconventional biexciton states in quantum dots. *Nature communications*, *5*(1), p.5721.
49. Barettin, D., Pecchia, A., der Maur, M.A., Di Carlo, A., Lassen, B. and Willatzen, M., 2021. Electromechanical field effects in InAs/GaAs quantum dots based on continuum $\vec{k}\cdot\vec{p}$ and atomistic tight-binding methods. *Computational Materials Science*, *197*, p.110678.
50. Guyot-Sionnest, P., 2012. Electrical transport in colloidal quantum dot films. The Journal of Physical Chemistry Letters, 3(9), pp.1169-1175.
51. Jung, S.M., Lee, T.H., Bang, S.Y., Han, S.D., Shin, D.W., Lee, S., Choi, H.W., Suh, Y.H., Fan, X.B., Jo, J.W. and Zhan, S., 2021. Modelling charge transport and electro-optical characteristics of quantum dot light-emitting diodes. *npj Computational Materials*, *7*(1), p.122.

52. Trellakis, A., Andlauer, T. and Vogl, P., 2005, June. Efficient solution of the schrödinger-poisson equations in semiconductor device simulations. In *International Conference on Large-Scale Scientific Computing* (pp. 602-609). Berlin, Heidelberg: Springer Berlin Heidelberg.

53. Trellakis, A., Zibold, T., Andlauer, T., Birner, S., Smith, R.K., Morschl, R. and Vogl, P., 2006. The 3D nanometer device project nextnano: Concepts, methods, results. *Journal of Computational Electronics*, *5*, pp.285-289.

54. Kumar, B., Campbell, S.A. and Paul Ruden, P., 2013. Modeling charge transport in quantum dot light emitting devices with NiO and ZnO transport layers and Si quantum dots. *Journal of Applied Physics*, *114*(4).

55. Verma, U.K. and Kumar, B., 2017. Charge transport in quantum dot organic solar cells with Si quantum dots sandwiched between poly (3-hexylthiophene)(P3HT) absorber and bathocuproine (BCP) transport layers. *Journal of Applied Physics*, *122*(15).

56. Jo, J.W., Kim, Y., Hou, B., Jung, S.M. and Kim, J.M., 2024. Charge transport dynamics and emission response in quantum-dot light-emitting diodes for next-generation high-speed displays. *Materials Today Physics*, *46*, p.101492.

57. Samarakoon, C., Choi, H.W., Lee, S., Fan, X.B., Shin, D.W., Bang, S.Y., Jo, J.W., Ni, L., Yang, J., Kim, Y. and Jung, S.M., 2022. Optoelectronic system and device integration for quantum-dot light-emitting diode white lighting with computational design framework. *Nature Communications*, *13*(1), p.4189.

58. Tewordt, M., Hughes, R.J.F., Martin-Moreno, L., Nicholls, J.T., Asahi, H., Kelly, M.J., Law, V.J., Ritchie, D.A., Frost, J.E.F., Jones, G.A.C. and Pepper, M., 1994. Vertical tunneling between two quantum dots in a transverse magnetic field. *Physical Review B*, *49*(12), p.8071.

59. Payne, M.C., 1986. Transfer Hamiltonian description of resonant tunnelling. *Journal of Physics C: Solid State Physics*, *19*(8), p.1145.

60. W. Van Roosbroeck, W. Shockley, "Photon-radiative recombination of electrons and holes in germanium," *Physical Review* 94(6) 1558 (1954).

61. R. Bhattacharya, B. Pal, B. Bansal, "On conversion of luminescence into absorption and the van Roosbroeck-Shockley relation," *Applied Physics Letters* 100(22) 222103 (2012).

62. Z. Hens and I. Moreels, Light absorption by colloidal semiconductor quantum dots, *J. Mater. Chem.* 22 (2012) 10406-10415. https://doi.org/10.1039/C2JM30760J.
63. G.A. Mantashian, P.A. Mantashyan, H.A. Sarkisyan, E.M. Kazaryan, G. Bester, S. Baskoutas, D.B. Hayrapetyan, Exciton-related Raman scattering, interband absorption and photoluminescence in colloidal CdSe/CdS core/shell quantum dots ensemble. *Nanomaterials*, 11(5), p.1274 (2021).
64. G.A. Mantashian, N.A. Zaqaryan, P.A. Mantashyan, H.A. Sarkisyan, S. Baskoutas, D.B. Hayrapetyan, Linear and nonlinear optical absorption of cdse/cds core/shell quantum dots in the presence of donor impurity. *Atoms*, 9(4), p.75 (2021).